\documentclass{ws-ijmpa}
\usepackage{epsfig}
\usepackage{fancybox}
\usepackage{euscript}
\usepackage{amsmath}
\usepackage{amssymb}
\usepackage{color}
\usepackage{cite}

\def\mathswitchr#1{\relax\ifmmode{\mathrm{#1}}\else$\mathrm{#1}$\fi}

\newcommand {\pslash}{\hbox{$\not\hbox{\kern-2.3pt $p$}$}}

\def\WAS{W\c AS}

\def\alf1{ {\alpha\over\pi} }

\begin{document}

\markboth{B.F.L. WARD, C. GLOSSER, S. JADACH, W. PLACZEK, M. SKRZYPEK, Z. \WAS,
S.A. YOST}
{NEW RESULTS ON PRECISION STUDIES OF HEAVY VECTOR BOSON PHYSICS}

%
\catchline{}{}{}{}{}
%

\title{NEW RESULTS ON PRECISION STUDIES OF HEAVY VECTOR BOSON PHYSICS}

\author{\footnotesize B.F.L. WARD}

\address{Department of Physics, Baylor University, One Bear Place \#97316\\
Waco, Texas 76798-7316,
USA
}

\author{\footnotesize C. GLOSSER}

\address{Department of Physics, Southern Illinois University, Box 1654\\
 Edwardsville, IL 62026-1654,  USA
}

\author{\footnotesize S. JADACH}

\address{Institute of Nuclear Physics, ul. Radzikowskiego 152\\
31-342 Krak\'ow, Poland\\
Theory Division, CERN\\
CH-1211 Geneva 23, Switzerland
}

\author{\footnotesize W. PLACZEK}

\address{Institute of Physics, Jagiellonian University, ul. Reymonta 4\\
30-059 Krak\'ow, Poland}

\author{\footnotesize M. SKRZYPEK}

\address{Institute of Nuclear Physics, ul. Radzikowskiego 152\\
31-342 Krak\'ow, Poland
}

\author{\footnotesize Z. \WAS}

\address{Institute of Nuclear Physics, ul. Radzikowskiego 152\\
31-342 Krak\'ow, Poland
}

\author{\footnotesize S. A. YOST}

\address{Department of Physics, Baylor University, One Bear Place \#97316\\
 Waco, Texas 76798-7316, USA}

\maketitle


\begin{abstract}
We present new results for two important heavy vector boson
physics processes: (1), virtual corrections to hard bremsstrahlung
which are relevant to precision predictions for the radiative
return process in Z boson production at and beyond LEP2 energies
; and, (2), electric charge screening effects in single W production
with finite $p_T$, multiple photon radiation in high energy collider
physics processes. In both cases we show that we improve the respective
precision tag significantly. Phenomenological implications are
discussed.
\end{abstract}

\section{Introduction}
Electroweak(EW)~\cite{sm1} and QCD~\cite{sm2} 
loop corrections are established:
precision LEP~\cite{lewwg} physics, $m_t$~\cite{fnal1}, $\ldots$, set a stage
for 1 GeV - 1 TeV high precision Standard Model~\cite{sm1,sm2}
tests via theoretical predictions for both signal and background
processes in high energy colliding beam environments. In the EW sector, this
now requires exact ${\cal O}(\alpha^2)$, ${\cal O}(\alpha^3L^3)$,
where $L$ is the respective big log, on an event-by-event basis in such
studies as radiative return from 1-2 GeV to the 
$\pi\pi$ resonance regime in  Daphne and the asymmetric B-Factories,
radiative return from 200 GeV to the Z pole in final LEP2 data analysis,
Z factory physics at ILC, ... .

In this paper, we present new results on two aspects of such
precision studies: (1), the virtual correction to $1\gamma$-bremsstrahlung;
(2), electric charge screening in $1W$ production~\cite{snglW} -- see also Ref.~\cite{passarino} in this connection.
\section{Virtual Corrections to Hard Bremsstrahlung}
\begin{figure}[!ht]
\begin{center}
\setlength{\unitlength}{1mm}
\begin{picture}(100,60)
\put(0, -1){\makebox(0,0)[lb]{
\epsfig{file=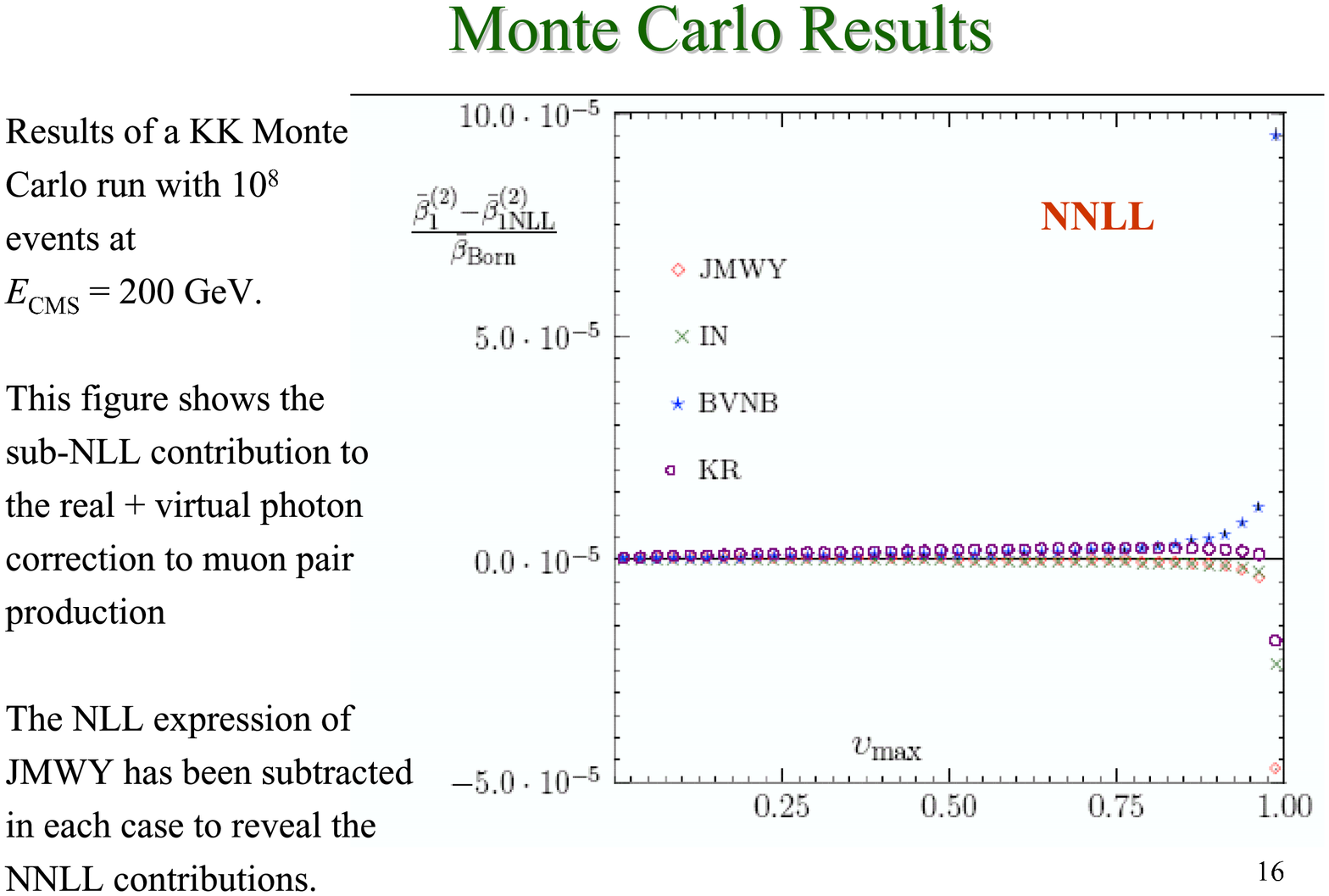,width=80mm,height=62mm,angle=-90}
}}
\end{picture}
\end{center}
\caption{Sub-NLL contribution $\beta_1^{(2)} - \beta^{(2)}_{1 \rm NLL}$.}
\label{fig:Figs-BNNLL}
\end{figure}

For the process $e^+e^-\rightarrow \bar{f} f+\gamma$,
we compare in Fig.~\ref{fig:Figs-BNNLL} the calculations in
Refs.~\cite{IN,BER,JMWY,KR} at the $\bar{\beta_1}^{(2)}$ level for
initial state radiation,
where $\{\bar\beta_n\}$ are the standard YFS~\cite{yfs} residuals.
The result by Ref.~\cite{IN}, labeled IN in the figure, is exact
and fully differential but without complete mass corrections,
the result in Ref.~\cite{BER}, labeled BVNB, is exact with the
complete mass corrections but is integrated over the photon
azimuthal angle, the result of Ref.~\cite{JMWY}, labeled JMWY,
is fully differential
with the complete mass corrections following the method of
Ref.~\cite{masscorr} whereas the exact result of Ref.~\cite{KR}, labeled KR,
is also fully differential with complete mass corrections
included in an entirely different way from that used in Ref.~\cite{JMWY}.
The agreement shown in the figure is at the $3\times 10^{-5}$ level in units of the
Born $e^+e^-\rightarrow \bar{f}f$ cross section for an energy cut at
$v_{max}=0.9625$.

\section{Electric Charge Screening Effects in $1W$ Production}

\begin{figure}[!ht]
\setlength{\unitlength}{0.1mm}
\hskip 1cm
\epsfig{file=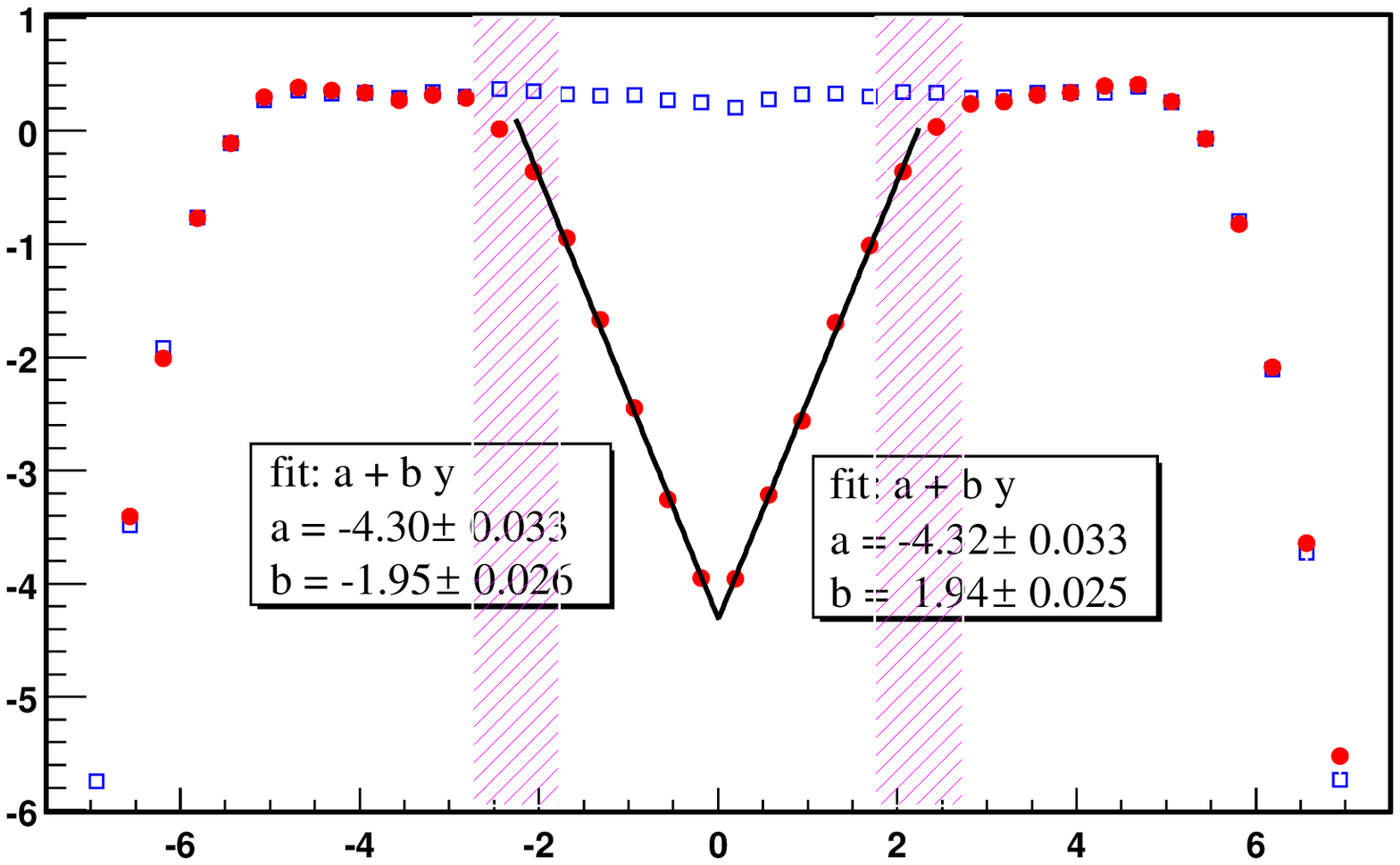,width=100mm,height=62mm}
\caption{\small\sf
  $\log_{10}$ of 
  $d\sigma/d\log_{10}\tan(\theta_{\gamma}/2)$ 
  with (red dots) and without (blue open squares) the
  ECS correction, arbitrary units. In boxes the values of fits are
  shown.
}
\begin{picture}(1000,0)
\put(625,119){\large $y=-\log_{10}\tan(\theta_{\gamma}/2)$}
\put(0,600){\large $\log_{10}\frac{d\sigma}{dy}$}
\put(0,550){\large  [arb.~units]}
\end{picture}
\label{fig:Fig1}
\end{figure}

Electric charge screening(ECS)/Leading Log scale transmutation(LLST)~\cite{snglW,passarino} is
known from low angle Bhabha scattering~\cite{yfsbhab} -- $L(s)\equiv ln\frac{s}{m_e^2}\Rightarrow L(|t|)$ in the LL expansion. 
In Ref.~\cite{snglW}, we have found
in the toy model
\begin{equation}
  \mu^-(p_a)+\mu^+(p_b)   \to  \mu^-(p_c)+\mu^+(p_d) +\gamma(k)
\end{equation}
the ECS corrected weight
\begin{equation}
  \tilde{S}_{ab}(k) W_{\rm ECS}(k)
\end{equation}
for the ISR IR emission factor $\tilde{S}_{ab}(k)$~\cite{yfs,yfsbhab}
where
\begin{equation}
  W_{\rm ECS}(k) = \frac{\tilde{S}_{abcd}(k)}{\tilde{S}_{ab}(k)+\tilde{S}_{cd}(k)},
  \label{eq:w_ecs}
\end{equation}
in a standard YFS notation.
For the single W production 
$e^-e^+\to f_c(p_c) +\bar{f}_d(p_d) +f_e(p_e) +\bar{f}_f(p_f) $
we find that we can do the same:{\small
\begin{equation}
 W_{\rm ECS}^{real} = \prod_{i} w^{R}(k_i), \;\;\;
 w^{R}(k)=\frac{
    \tilde{S}_{ab}(k) +\tilde{S}_{CD}(k) +\tilde{S}_{aC}(k) +\tilde{S}_{bD}(k) 
   +\tilde{S}_{aD}(k) +\tilde{S}_{bC}(k) }
 { \tilde{S}_{ab}(k) +\tilde{S}_{CD}(k) }.
\label{waga-real}
\end{equation}}\noindent
for the effective~\cite{snglW} final particles 'C' and 'D' close to the
incoming beams, as we illustrate in Fig.~\ref{fig:Fig1}.
A factor $\exp(\Delta U)$ cancels {\em exactly} the dummy~\cite{snglW} 
IR $\epsilon$-dependence
and compensates {\em approximately} for the normalization change due
to the $\langle W_{\rm ECS}^{real} \rangle$ weight and
the effective coupling is also that at $|t|$, by standard 
renormalization group~\cite{bflwyfs} arguments; this all
%
is realized~\cite{snglW} with the normalization correction(here,$\gamma_r\equiv \frac{2\alpha}{\pi}(L(|r|)-1),\bar\gamma_t\equiv\frac{1}{2}(\gamma_t+\gamma_s)$){\small
\begin{equation}
\begin{split}
& W_{ECS}^{norm} = \exp \left(\frac{3}{4}(\bar\gamma_t -\gamma_s)\right) 
                   \exp \big(\Delta U(\epsilon)\big) 
\\
& \Delta U(\epsilon)= U(\epsilon) -U_R(\epsilon), \;\;
U(\epsilon)= \int\limits_{\epsilon \sqrt{s}/2}^{\sqrt{s}} \frac{d^3k}{k^0}
                      \tilde{S}_{ab}(k), \;\;
U_R(\epsilon)= \int\limits_{\epsilon \sqrt{s}/2}^{\sqrt{s}} \frac{d^3k}{k^0}
                      \tilde{S}_{ab}(k)w^R(k).
\label{waga-vs}
\end{split}
\end{equation}}\noindent
to maintain the exact IR cancellation in the MC (KoralW~\cite{snglW,krlw}, for example).\par

The only purpose of the weight $W_{\rm ECS}^{real}$ is to restore the ECS effect due to  ISR$\otimes$FSR interference. We do not aim at re-creating the FSR. This would be formally possible with a similar weight; however, the resultant
weight distribution would be bad and the attendant MC calculation
would not be convergent.
We get $W_{\rm ECS}^{real}\to 1$ for photons collinear with the FS effective fermions $C$ and $D$. This ensures a very good weight distribution.
The FSR can be treated separately, either inclusively (calorimetric acceptance)
or exclusively, generated with the help of PHOTOS~\cite{photos}
\footnote{Care has to be taken to implement ECS for FSR, if necessary.}.
The precision tag of $\le 2\%$ is realized~\cite{snglW} -- good enough for 
final LEP2 data analysis.

\section*{Acknowledgments}
Five of the authors (S.J.,W.P.,M.S.,B.F.L.W.,Z.W.) would like to thank 
Prof.\ G.\ Altarelli of the CERN TH Div.\ and Prof. D. Schlatter
and the ALEPH, DELPHI, L3 and OPAL Collaborations, respectively, 
for their support and hospitality while part of this work was completed.
B.F.L.W.\ would like to thank Prof.\ C.\ Prescott of Group A at SLAC for his 
kind hospitality while part of this work was in its developmental stages.
Work supported in part by US DoE contract DE-FG05-91ER40627, by NATO grants
PST.CLG.97751,980342, by Polish Government grants KBN 5P03B09320,
2P03B00122, by European  Commission 5-th framework contract HPRN-CT-2000-00149 and by the Polish-French Collaboration within IN2P3 through LAPP Annecy.

\end{document}